\documentclass[twocolumn,amsmath,amssymb,aps,pra,longbibliography,floatfix]{revtex4-1}
\usepackage{physics}
\usepackage{amsmath}
\usepackage{cases}
\usepackage{url}
\usepackage{natbib}
\usepackage{textcase}
\usepackage{amssymb}
\usepackage{graphicx}
\usepackage{bm} 

\usepackage{times}
\usepackage{float}
\usepackage{multirow,microtype,color,relsize}
\usepackage{subfigure}
\usepackage[breaklinks=true]{hyperref}
\usepackage[utf8]{inputenc}
\usepackage[english]{babel}
\hypersetup{colorlinks=true,linkcolor=blue,citecolor=blue}
\hypersetup{linktocpage}
\usepackage{CJKutf8}
\usepackage{breakcites}
\usepackage{float}
\usepackage[dvipsnames]{xcolor}
\definecolor{mypine}{RGB}{1, 121, 111}

\def \be {\begin{equation}}
\def \ee {\end{equation}}

\begin{document}
\begin{CJK*}{UTF8}{gbsn}
\title{Plasmons in semiconductor and topological insulator wires with large dielectric constant}

\author{Yi Huang~(黄奕)}
\affiliation{School of Physics and Astronomy, University of Minnesota, Minneapolis, Minnesota 55455, USA}
\email[Corresponding author: ]{huan1756@umn.edu}

\author{Chao-Hsiang Sheu}
\affiliation{School of Physics and Astronomy, University of Minnesota, Minneapolis, Minnesota 55455, USA}

\author{B.\,I. Shklovskii}
\affiliation{School of Physics and Astronomy, University of Minnesota, Minneapolis, Minnesota 55455, USA}

\date{\today}

\begin{abstract}
The dispersion law of plasmons running along thin wires with radius $a$ is known to be practically linear. We show that in wires with a dielectric constant $\kappa$ much larger than that of its environment $\kappa_e$, such dispersion law crosses over to a dispersionless three-dimensional-like law when the plasmon wavelength becomes shorter than the length $(a/2) \sqrt{(\kappa/\kappa_e)\ln(\kappa/2\kappa_e)}$ at which the electric field lines of a point charge exit from the wire to the environment. This happens both in trivial semiconductor wires and wires of three-dimensional topological insulators.

\end{abstract}
\maketitle
\end{CJK*}

We are honored to contribute to the issue dedicated to Mark Azbel's 90 birthday. Mark was a brilliant physicist who has made classical contributions to physics of metals. He also had enormously broad interests in science beyond physics. When one of us was writing the paper on derivation of the Gompertz law of human mortality, Mark's advice was indispensable.

\section{Introduction}
\label{sec:intro}
Plasma waves or plasmons in electron gas in semiconductors continue to attract an enormous attention.
Plasmon dispersion $\omega_p(q)$ depends on the dimensionality of the system. 
For three-dimensional (3D) electron gas with parabolic energy spectrum
$\varepsilon_{\vb{k}} =\hbar^2 k^2/2m^{\star}$, plasmons are practically dispersionless, i.e.
\begin{align}\label{eq:omega_3d0}
    \omega_{p3} = \sqrt{\frac{4\pi e^2 n_3}{\kappa m^{\star}}}.
\end{align}
This law is valid in the long wavelength limit $q \ll \sqrt{k_F/a_B}$, where $k_F$ is the Fermi wave vector defined through the Fermi level $\mu = \hbar^2 k_F^2/2m^{\star}$, and $a_B = \kappa \hbar^2/m^{\star} e^2$ is the effective Bohr radius. 
Here $n_3 = k_F^3/3\pi^2$ is the 3D concentration of electrons, $e$ is the elementary charge unit, $\kappa$ is the dielectric constant, and $m^{\star}$ is the effective mass of electrons. 

For thin films of thickness $d$ which can be considered as two-dimensional (2D) systems at $k_F d \ll 1$, the plasmon dispersion at $q \ll d^{-1}$ reads~\cite{stern1967,chaplik1980,ando1982,chaplik1985}
\begin{align} \label{eq:omega_2d0}
    \omega_{p2}(q) = \sqrt{\frac{2e^2 \mu q}{\kappa \hbar^2}} = \sqrt{\frac{2\pi e^2 n_2 q}{\kappa m^{\star}}},
\end{align}
where the 2D concentration of electrons is given by $n_2 = k_F^2/2\pi$.

For thin cylindrical wires with radius $a$ which can be considered as one-dimensional (1D) systems at $k_F a \ll 1$, the plasmon dispersion law at $q \ll a^{-1}$ is given by~\cite{sommerfeld1899,chaplik1980,Friesen:1980,li_das_1991}
\begin{align}\label{eq:omega_1d0}
    \omega_{p1}(q) = \sqrt{\frac{2e^2 n_1q^2 \ln (1/qa)}{\kappa m^{\star}}}.
\end{align}
where $n_1 = 2k_F/\pi$ is the 1D concentration of electrons.

\begin{figure}[t]
    \centering
    \includegraphics[width = \linewidth]{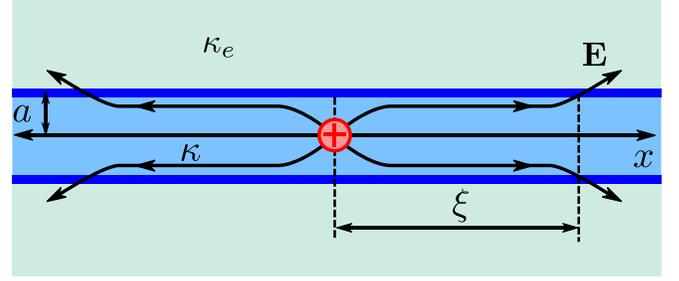}
    \caption{Schematic plot of a wire (blue) of radius $a$ and dielectric constant $\kappa$. The dielectric constant for the environment (green) is $\kappa_e$. Electric field lines (black) start at the point charge shown by a red circle and channel through the wire until exiting outside at $x\sim \xi \gg a$. Here $\xi \approx (a/2) \sqrt{(\kappa/\kappa_e)\ln(\kappa/2\kappa_e)} \gg a$ is the characteristic length much larger than $a$ if $\kappa \gg \kappa_e$~\cite{keldysh1997,finkelstein2002,teber2005,kamenev2006,cui2006,huang2021c}.}
    \label{fig:wire}
\end{figure}

Above we are taking about the cases when the dielectric constant $\kappa$ of the film is equal or comparable to the dielectric constant of its environment $\kappa_e$. 
However, there are many films with huge dielectric constant $\kappa \gg \kappa_e$. 
For example, $\kappa$ can be as large as $1000$ for PbTe or $20000$ for STO~\cite{huang2021b}. 
Moreover, intensively studied 3D topological insulators (TIs) compounds based on Bi$_2$Te$_3$ also have large dielectric constants $\kappa \sim 200$~\cite{richter1977,borgwardt2016,bomerich2017}. 
Such films are often deposited on substrates with a much smaller dielectric constant $\kappa \sim 4$ like silicon oxide or hexagonal boron nitride (hBN). 
It was shown recently that such enormous dielectric-constant contrast strongly modifies the plasmon dispersion both in trivial semiconductor films~\cite{bondarev2017} and for TI films~\cite{Di_Pietro:2013,stauber2013}.

\begin{figure}[t]
    \centering
    \includegraphics[width = \linewidth]{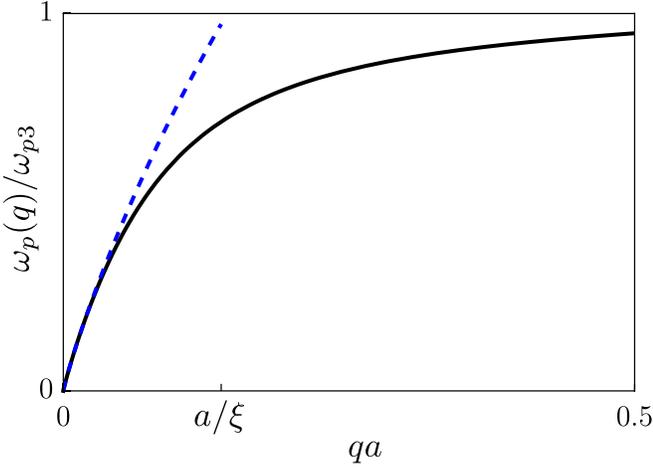}
    \caption{Plasmon dispersion along the wire $\omega_{p}(q)$ following by Eqs.~\eqref{eq:omega_trivial} or \eqref{eq:omega_ti} with angular momentum $m=0$, in units of 3D plasma frequency $\omega_{p3}$ at large $\kappa/\kappa_e$. We use $\kappa/\kappa_e = 50$ in this plot, so that $\xi \simeq 7 a$ given by Eq.~\eqref{eq:xi}. For trivial wires or TI wires at $k_F a\ll 1$, $\omega_{p3}$ is given by Eq.~\eqref{eq:omega_3d0}. For TI wires at $k_F a \gg 1$, $\omega_{p3}$ follows Eq.~\eqref{eq:omega_ti_3d}. At $a/\xi < qa < 1$, the plasmon frequency is practically independent on $q$ and equal to $\omega_{p3}$. At $qa \ll a/\xi$, the plasmon frequency $\omega_{p}(q) \propto q [\ln (qa)^{-1}]^{1/2}$ (dashed blue curve) follows Eq.~\eqref{eq:omega_1d0} for trivial wires and TI wires at $k_F a\ll 1$, while for TI wires at $k_F a\gg 1$, $\omega_p(q)$ follows Eq.~\eqref{eq:omega_ti_1d}. }
    \label{fig:omega}
\end{figure}

In this paper we study the role of a strong dielectric-constant contrast $\kappa \gg \kappa_e$ for the plasmon dispersion in thin wires of a trivial semiconductor and of a 3D TI. We show that at wave vectors $\xi^{-1} < q < a^{-1}$, the confinement of electric fields of an electron inside the wire makes the plasmon dispersionless similarly to 3D result Eq.~\eqref{eq:omega_3d0}. 
Here 
\begin{align}\label{eq:xi}
    \xi \approx (a/2) \sqrt{(\kappa/\kappa_e)\ln(\kappa/2\kappa_e)}  \gg a
\end{align}
is the characteristic length at which the electric field of a point charge in the wire exits to the environment (see Fig.~\ref{fig:wire})~\cite{keldysh1997,finkelstein2002,teber2005,kamenev2006,cui2006,huang2021c}. 
On the other hand, at $q \ll \xi^{-1}$, the large dielectric constant $\kappa$ of the wire plays no role, because after exiting from the wire the electric field lines are mostly located in the environment, and the electrostatics is determined by $\kappa_e$. 
As a result, the plasma frequency $\omega_{p}$ at $q\xi \ll 1$ has practically linear dispersion similar to Eq.~\eqref{eq:omega_1d0} with $\kappa$ replaced by $\kappa_e$. Predicted crossover from the practically linear dispersion to the dispersionless plateau with growing $q\xi$ is illustrated by Fig.~\ref{fig:omega}. 

The plan of this paper is as follows.
In Sec.~\ref{sec:trivial} we derive the plasmon dispersion for trivial semiconductor wires in both cases $\kappa = \kappa_e$ and $\kappa \gg \kappa_e$.
The results for both the thick wire limit $k_F a \gg 1$ and thin wire limit $k_F a \ll 1$ are also discussed in Sec.~\ref{sec:trivial}.
In Sec.~\ref{sec:ti} we do the similar analysis for TI wires. 
We conclude in Sec.~\ref{sec:conclusion}.

\section{Trivial wires}
\label{sec:trivial}
Consider a conventional semiconductor wire of radius $a$, length $L$, and dielectric constant $\kappa$, in an environment with dielectric constant $\kappa_e$. Below we calculate the plasmon dispersion of this wire at zero temperature.

The plasmon frequency $\omega_p (\vb{q})$ is determined by zeros of the dielectric function $\epsilon(\vb{q}, \omega_p) = 0$.
In random phase approximation (RPA), the dielectric function $\epsilon(\vb{q},\omega)$ can be written as 
\begin{align}\label{eq:epsilon}
    \epsilon(\vb{q}, \omega) = 1 - v(\vb{q}) \Re \Pi(\vb{q}, \omega) 
\end{align}
where $v(\vb{q})$ is given by the matrix element of Coulomb interaction between electrons and $\Re \Pi(\vb{q}, \omega)$ is the real part of the polarization bubble.
The RPA polarization bubble at zero temperature is given by
\begin{align}\label{eq:pi}
    \Pi(\vb{q}, \omega) = \frac{g_s}{V}\sum_{\vb{k}} \frac{\Theta(\mu - \varepsilon_{\vb{k}}) - \Theta(\mu - \varepsilon_{\vb{k} + \vb{q}})}{\varepsilon_{\vb{k}} - \varepsilon_{\vb{k} + \vb{q}} + \hbar \omega + i \delta},
\end{align}
where $g_s = 2$ is the spin degeneracy, $V = \pi a^2 L$ is the total volume of the wire, $\mu$ is the Fermi level, and $\Theta(x)$ is the Heaviside theta function. $\vb{k} = (k_x, k_{\phi}, k_{\rho})$ is the 3D momentum vector conjugated to the position vector in cylindrical coordinates $\vb{r} = (x, \phi, \rho)$. 
The energy dispersion for an electron in a conventional semiconductor wire reads 
\begin{align}\label{eq:energy_nr}
    \varepsilon(\vb{k}) = \frac{\hbar^2 (k_x^2 + k_{\phi}^2 + k_{\rho}^2)}{2 m^{\star}}
\end{align}
where $m^{\star}$ is the effective mass in the semiconductor.
In general when $a \ll L$ one can treat $k_x$ as a continuous variable but $k_{\phi}$ and $k_{\rho}$ are discrete because of the quantization in transverse direction along $\phi$ and $\rho$. As a result, the energy dispersion $\varepsilon(\vb{k})$ consists of discrete subbands labeled by $k_{\phi}$ and $k_{\rho}$. 

Let us first calculate the polarization bubble in the limit when the Fermi level is much larger than the subband gap, $\mu \gg \hbar v/a$ or $k_F a \gg 1$, where $k_F = \sqrt{2 m^{\star} \mu/\hbar^2}$.
In this case one can ignore the quantization and treat the whole momentum vector $\vb{k}$ as a continuous variable. In this case, the real part of the polarization bubble is given by the well-known 3D Lindhard function
\begin{align}\label{eq:pi_3d0}
    \Re \Pi(q, \omega) &= \frac{ 2 m^{\star} k_F}{(2\pi \hbar)^2} \left[-1 + \frac{k_F}{2 q}(1- y_{-}^2) \ln\abs{\frac{1-y_{-}}{1+y_{-}}} \right.\nonumber \\ 
    &\left.- \frac{k_F}{2 q}(1- y_{+}^2) \ln\abs{\frac{1+y_{+}}{1-y_{+}}}\right]
\end{align}
where $y_{\pm} = \frac{\hbar \omega k_F}{2\mu q} \pm \frac{q}{2k_F}$.

Consider the long wave length limit $q \to 0$ with fixed $\omega$ in the interval $(\hbar q^2/2m^{\star}, \mu/\hbar)$, we can simplify Eq.~\eqref{eq:pi_3d0} to the leading order of $q$
\begin{align}\label{eq:pi_3d1}
    \Re \Pi(q, \omega) = \frac{ 2 m^{\star} k_F}{(2\pi \hbar)^2} \frac{8(q/k_F)^2}{3(\hbar\omega/\mu)^2} = \frac{n_3 q^2}{m^{\star} \omega^2},
\end{align}
where in the last step we use the 3D concentration of electrons $n_3 = k_F^3/3\pi^2$.
In the case of $\kappa = \kappa_e$, the Coulomb interaction is given by the conventional form 
\begin{align}\label{eq:vq_3d}
    v(q) = 4\pi e^2/\kappa q^2.
\end{align}
Substituting Eqs.~\eqref{eq:pi_3d1} and \eqref{eq:vq_3d} into Eq.~\eqref{eq:epsilon} and solving equation $\epsilon(q,\omega_p) = 0$ for $\omega_p$, we obtain the familiar 3D plasma frequency as shown in Eq.~\eqref{eq:omega_3d0}.

In the opposite limit such that the Fermi level is much smaller than the subband gap $\mu \ll \hbar v/a$ or $k_F a \ll 1$, only the first subband with dispersion $\varepsilon(k_x) = \hbar^2k_x^2/2m^{\star} $ contributes to the polarization bubble.
The polarization bubble is given by the 1D Lindhard function
\begin{align}\label{eq:pi_1d0}
    \Re \Pi(q, \omega) = \frac{1}{\pi a^2} \frac{m^{\star}}{\pi \hbar^2 q} \ln\abs{\frac{(\hbar \omega)^2 - [\tfrac{\hbar^2}{2m^{\star}}(2k_Fq + q^2)]^2}{(\hbar \omega)^2 - [\tfrac{\hbar^2}{2m^{\star}}(2k_Fq - q^2)]^2}}
\end{align}
Consider the long wave length limit $q \to 0$ with fixed $\omega$ in the interval $(\hbar q^2/2m^{\star}, \mu/\hbar)$, we can simplify Eq.~\eqref{eq:pi_1d0} to the leading order of $q$
\begin{align}\label{eq:pi_1d1}
    \Re \Pi(q, \omega) = \frac{2k_F q^2}{\pi^2 a^2 m^{\star} \omega^2} = \frac{n_1}{\pi a^2} \frac{q^2}{m^{\star} \omega^2},
\end{align}
where in the last step we use the 1D concentration $n_1 = 2k_F/\pi$ for the the case of a single subband.
If $\kappa = \kappa_e$, the matrix element of the conventional Coulomb interaction  $v(\vb{r}, \vb{r}') = e^2/\kappa \abs{\vb{r} - \vb{r}'}$ in two-particle energy eigenstates for the first subband reads
\begin{align}\label{eq:mel_v0}
    &\mel{p',k'}{v(\vb{r}, \vb{r}')}{p, k} \nonumber\\
    &= \frac{1}{V^2} \int d x \, d x'\,  e^{-i (p'-p) x - i(k'-k) x'} \nonumber \\
    &\times \int d^2r_{\perp} d^2r_{\perp}' \Psi(\vb{r}_{\perp})^2  \Psi(\vb{r}_{\perp}')^2 v(\vb{r}, \vb{r}')  \nonumber \\
    &= \frac{1}{V} \sum_q v(q) \delta_{-p'+q+p,0} \delta_{-k'-q+k,0},
\end{align}
where we assume the one-particle electron wavefunction takes the form $\psi_k(\vb{r}) = V^{-1/2} e^{ikx} \Psi(\vb{r}_{\perp})$, where $\vb{r}_{\perp}$ is the projection of the three-dimensional vector $\vb{r}$ in the plane perpendicular to the wire. Here $\Psi(\vb{r}_{\perp})$ is the first-subband wavefunction in the transverse plane. 
Because $\Psi(\vb{r}_{\perp})$ is confined within radius $a$, in the limit of $qa\ll1$ the expression of $v(q)$ in Eq.~\eqref{eq:mel_v0} to the leading order can be written as~\footnote{In the derivation of Eqs.~\eqref{eq:vq_1d} and \eqref{eq:omega_1d0}, we assume the electrons occupied only the first subband for convenience, but actually the 1D plasmon result is valid as long as $q \ll a^{-1}$ independent on the value of $k_F a$. This can be seen by solving the Laplace equation of the electric potential in classical electrostatics similarly to Eq. (2.2) in Ref.~\onlinecite{chaplik1985}.}
\begin{align}\label{eq:vq_1d}
    v(q) = 2 \pi a^2 (e^2/\kappa)\ln (1/qa).
\end{align}
Substituting Eq.~\eqref{eq:pi_1d1} and \eqref{eq:vq_1d} into Eq.~\eqref{eq:epsilon} and solving equation $\epsilon(q,\omega_p) = 0$ for $\omega_p$, we obtain the 1D plasma frequency given by Eq.~\eqref{eq:omega_1d0}.

However, in the case of $\kappa \neq \kappa_e$, the expression of Coulomb interaction will be modified by a different electrostatics.
For an electron sitting at a source point $\vb{r}' = (x',\phi',\rho')$ with $\rho' < a$, the Coulomb interaction with the other electron at a field point $\vb{r} = (x,\phi, \rho)$ with $\rho < a$ is given by~\cite{cui2006}
\begin{align}\label{eq:v2}
    &v(\vb{r}, \vb{r}') = \frac{4e^2}{\pi \kappa} \int_0^{\infty} dk \cos[k(x-x')] \left[\frac{1}{2} I_0(k \rho_{<}) K_0(k \rho_{>}) \right. \nonumber \\ 
    &+ \sum_{m=1}^{\infty} I_{m}(k \rho_{<}) K_m(k \rho_{>}) \cos[m (\phi - \phi')] \nonumber \\ 
    & - \frac{1}{2} \frac{f_0(ka) K_0(ka)}{I_0(ka)} I_0(k \rho_{<}) I_0(k \rho_{>})  \nonumber \\
    & - \left.\sum_{m=1}^{\infty} \frac{f_m(ka) K_m(k a)}{I_m(k a)} I_{m}(k \rho_{<}) I_m(k \rho_{>}) \cos[m (\phi - \phi')]\right]
\end{align}
where $\rho_{<}$ ($\rho_{>}$) indicates the smaller (larger) radial coordinates of the source and the field point. $I_m(z)$ and $K_m(z)$ are modified Bessel function of order $m$. The function $f_m(ka)$ is given by
\begin{align}\label{f_m}
    f_m(ka) = \frac{1 - \kappa/\kappa_e}{1 - g_m(ka) \kappa/\kappa_e}, 
\end{align}
where 
\begin{align}\label{g_m}
    g_m(ka) = \frac{K_m(ka) I_m'(ka)}{K_m'(ka) I_m(ka)}.
\end{align}
Next we calculate $v(\vb{q})$ corresponding to Eq.~\eqref{eq:v2}. 
The matrix element of $v(\vb{r}, \vb{r}')$ in two-particle energy eigenstates reads
\begin{align}\label{eq:mel_v1}
    &\mel{\vb{p}',\vb{k}'}{v(\vb{r}, \vb{r}')}{\vb{p}, \vb{k}} \nonumber\\
    &= \frac{1}{(2\pi a L)^2} \int d^3 r \, d^3 r'\,  e^{-i p_x' x - ik_x' x' + ik_x x' + i p_x x} \nonumber \\
    &\times e^{-i p_{\phi}' \phi - ik_{\phi}' \phi' + ik_{\phi} \phi' + i p_{\phi} \phi} \mel{p_{\rho}', k_{\rho}'}{v(\vb{r}, \vb{r}')}{p_{\rho}, k_{\rho}} \nonumber \\
    &\approx \frac{1}{V}  \sum_{\vb{q}} v(q,m) \delta_{-\vb{p}'+\vb{q}+\vb{p},0} \delta_{-\vb{k}'-\vb{q}+\vb{k},0} \delta_{p_{\rho}', p_{\rho}}\delta_{k_{\rho}', k_{\rho}},
\end{align}
where in the second step $\ket{p_{\rho}, k_{\rho}}$ represents the eigenstate in the radial direction.
In the last step we use the fact that $v(\vb{r}, \vb{r}')$ is almost independent on the radial coordinates $\rho$ and $\rho'$ if the longitudinal distance $\abs{x - x'} > a$, such that
\begin{align}
    \mel{p_{\rho}', k_{\rho}'}{v(\vb{r}, \vb{r}')}{p_{\rho}, k_{\rho}} \approx \qty(\frac{2}{a})^2 v(\vb{r}, \vb{r}') \delta_{p_{\rho}', p_{\rho}}\delta_{k_{\rho}', k_{\rho}}.
\end{align}
Substituting Eq.~\eqref{eq:v2} into Eq.~\eqref{eq:mel_v1} with $\rho = \rho' = a$, we arrive at 
\begin{align}\label{eq:vqm}
    v(q,m) = \frac{2\pi e^2 a^2}{\kappa} [1-f_{\abs{m}}(qa)] I_{\abs{m}}(qa) K_{\abs{m}}(qa)
\end{align}
Eq.~\eqref{eq:vqm} is justified as long as $qa < 1$.

Interestingly, for both cases $k_F a \gg 1$ and $k_F a \ll 1$, the corresponding plasmon dispersion has the same form if $q < 1/a$.
In the case $k_F a \gg 1$, substituting Eq.~\eqref{eq:pi_3d1} and \eqref{eq:vqm} into Eq.~\eqref{eq:epsilon} and solving equation $\epsilon(q, \omega_p) = 0$ for $\omega_p$, we arrive at the plasmon dispersion 
\begin{align}
    \omega_p^2(q,m) = &\frac{2 e^2 n_1 (q^2a^2 + m^2) }{\kappa m^{\star} a^2}\nonumber\\
    & \times [1-f_{\abs{m}}(qa)] I_{\abs{m}}(qa) K_{\abs{m}}(qa), \label{eq:omega_trivial}
\end{align}
where $n_1 = n_3 \pi a^2$ is the effective 1D concentration of electrons.
For the other case $k_F a \ll 1$, replacing Eq.~\eqref{eq:pi_3d1} by \eqref{eq:pi_1d1} we obtain the same expression as Eq.~\eqref{eq:omega_trivial}, with $m=0$ and a modified definition of $n_1 = 2k_F/\pi$.
Eq.~\eqref{eq:omega_trivial} shows the plasma frequency is proportional to the square root of the electron concentration $\omega_p \propto \sqrt{n_1}$.

If $\kappa \gg \kappa_e$ Eqs.~\eqref{eq:vqm} and \eqref{eq:omega_trivial} at $m = 0$ can be further simplified as
\begin{align}\label{eq:vq0}
    v(q) \equiv v(q,0) \approx \frac{4\pi e^2}{\kappa(q^2 + \xi^{-2})},\; q< a^{-1},
\end{align}
\begin{align}\label{eq:omega_trivial0}
    \omega_p^2(q) \equiv \omega_p^2(q,0) \approx \frac{4e^2 n_1}{\kappa m^{\star} a^2} \frac{1}{1 + (q\xi)^{-2}},
\end{align}
where $\xi$ is the length determined by the equation $\xi^2 = a^2(\kappa/2\kappa_e)\ln(\xi/a)$ ~\cite{keldysh1997,finkelstein2002,teber2005,kamenev2006,cui2006,huang2021c}.
If $\kappa \gg \kappa_e$, to the leading order of $\kappa/\kappa_e$ we have $\xi \gg a$ given by Eq.~\eqref{eq:xi}.
From Eqs.~\eqref{eq:vq0} and \eqref{eq:omega_trivial}, in the limit $q \xi \gg 1$ we immediately see that $v(q)$ and $\omega_p(q)$ are equal to 3D expressions Eqs.~\eqref{eq:vq_3d} and \eqref{eq:omega_3d0} respectively, where $n_3 =  n_1/\pi a^2$.
In the opposite limit $q \xi \ll 1$, $v(q)$ and $\omega_p(q)$ are equal to 1D expressions Eqs.~\eqref{eq:vq_1d} and \eqref{eq:omega_1d0} respectively with $\kappa$ replaced by $\kappa_e$. This shows that the electrostatics at length scale $a<x<\xi$ is similar to the conventional 3D electrostatics with uniform dielectric constant $\kappa$, while at length scales $x>\xi$ we return to the conventional 1D electrostatics with uniform dielectric constant $\kappa_e$.
The crossover of $\omega_p(q)$ between two limits $q \xi \ll 1$ and $q \xi \gg 1$ is shown in Fig.~\ref{fig:omega}.

Next we discuss the range of $q$ at which Eq.~\eqref{eq:omega_trivial0} is valid.
Since the first correction to Eqs.~\eqref{eq:pi_3d1} and \eqref{eq:pi_1d1} is smaller than the leading term by a factor $(q\mu/ \hbar k_F \omega)^2$, this correction can be neglected if 
\begin{align}\label{eq:correction0}
    q\mu/ \hbar k_F \omega_p(q) < 1,
\end{align}
Combining with the other restriction $qa < 1$, Eq.~\eqref{eq:correction0} at $k_F a_B >1$ and $a_B > a$ leads to 
\begin{align}
    q <
\begin{cases}
a^{-1}, \, & k_F a > a_B/a, \\
 \sqrt{k_F/a_B}, \, & 1<k_F a < a_B/a, \\
a^{-1} (k_F a_B)^{-1/2},\, & k_F a < 1.
\end{cases}
\end{align}
If at $k_F a_B >1$ and $a_B < a$ then Eq.~\eqref{eq:correction0} leads to $q a <1$.

\section{Topological insulator wires}
\label{sec:ti}
In this section we discuss the plasmon dispersion for the surface electrons in a 3D topological insulator (TI) wires.
The polarization bubble of a TI wire is given by
\begin{align}\label{eq:pi_ti}
    \Pi(\vb{q}, \omega) = &\frac{1}{V}\sum_{\vb{k},s,s'} \left[F_{ss'}(\vb{k},\vb{k} + \vb{q}) \right.\nonumber \\
    & \left.\times \frac{\Theta(\mu - \varepsilon_{s\vb{k}}) - \Theta(\mu - \varepsilon_{s'\vb{k} + \vb{q}})}{\varepsilon_{s\vb{k}} - \varepsilon_{s'\vb{k} + \vb{q}} + \hbar \omega + i \delta}\right],
\end{align}
where the energy is given by the Dirac dispersion
\begin{align}\label{eq:energy_dirac}
    \varepsilon_{s\vb{k}} = s\hbar v_F \abs{\vb{k}} = s \hbar v_F \sqrt{k^2 + (l/a)^2},
\end{align}
Here $v_F$ is the Fermi velocity, $s = + $ ($-$) indicates the electron (hole) band, and $l \in \mathbb{Z} + 1/2$ is the angular momentum quantum number~\footnote{$l$ is half integer because of the anti-periodic boundary condition resulted from the $\pi$ Berry phase in TI.}, such that the 2D momentum vector is given by $\vb{k} = (k, l/a)$.
The function $F_{ss'}(\vb{k},\vb{k} + \vb{q})$ in Eq.~\eqref{eq:pi_ti} is defined as 
\begin{align}
    F_{ss'}(\vb{k},\vb{k} + \vb{q}) = \frac{1}{2}\qty(1 + ss'\frac{ \vb{k} \vdot (\vb{k} + \vb{q})} {\abs{\vb{k}}\abs{\vb{k} + \vb{q}}} ).
\end{align}

Let us start with the limit of Fermi level much larger than the subband gap $\mu \gg \hbar v_F/ a$ or equivalently $k_F a \gg 1$, where $k_F = \mu/\hbar v_F$. 
In this case, one deals with relatively thick wires and the surface curvature can be ignored.
Namely, we can neglect the subband quantization and treat the 2D momentum vector $\vb{k}$ as a continuous variable in the calculation of the polarization bubble Eq.~\eqref{eq:pi_ti}.
The result of the polarization bubble in the long wavelength limit, i.e. $q \to 0$ with fixed $\omega$ in the interval $(v_F q, \mu/\hbar)$, is well known as in the similar calculations for graphene~\cite{shung1986,wunsch2006,barlas2007,hwang2007,kotov2012}
\begin{align}\label{eq:pi_ti1}
    \Re \Pi(q, \omega) = \frac{\mu q^2}{2\pi a \hbar^2 \omega^2} = \frac{v_F \sqrt{n_2} q^2}{\sqrt{\pi} a \hbar \omega^2} = \frac{v_F \sqrt{n_1} q^2}{\sqrt{2 a^3} \pi  \hbar \omega^2},
\end{align}
where $n_2 = k_F^2/4\pi = \mu^2/4\pi\hbar^2 v_F^2$ is the 2D concentration of surface electrons, and $n_1 = 2\pi a n_2 $ is the effective 1D concentration of surface electrons.

Substituting Eqs.~\eqref{eq:pi_ti1} and \eqref{eq:vqm} into Eq.~\eqref{eq:epsilon} and solving equation $\epsilon(q, \omega_p) = 0$ for $\omega_p$ we arrive at the plasmon dispersion
\begin{align}
    \omega_p^2(q,m) = &\frac{e^2 \mu (q^2a^2 + m^2) }{\kappa \hbar^2 a} \nonumber \\
    &\times [1-f_{\abs{m}}(qa)] I_{\abs{m}}(qa) K_{\abs{m}}(qa). \label{eq:omega_ti}
\end{align}
If we compare the result for TI wires Eq.~\eqref{eq:omega_ti} with the result for trivial wire Eq.~\eqref{eq:omega_trivial}, we find that both of them have the behavior $\omega_p \propto \sqrt{\mu}$. 
However, since $\mu \propto \sqrt{n_1}$ in TI wires while $\mu \propto n_1$ in trivial wires, this leads to different behavior of plasmon dispersion as a function of electron concentration. 
Namely, $\omega_p \propto n_1^{1/4}$ in TI wires while $\omega_p \propto n_1^{1/2}$ in trivial wires.

The dielectric constant of a typical TI wire is $\kappa \sim 200$~\cite{richter1977,borgwardt2016,bomerich2017}. If the environment is made by materials of small dielectric constant, for example, SiO$_2$ with $\kappa_e = 4$, then $\kappa \gg \kappa_e$, and Eq.~\eqref{eq:omega_ti} with $m = 0$ can be further simplified following the similar discussion in Sec.~\ref{sec:trivial}:
\begin{align}\label{eq:omega_ti0}
    \omega_p^2(q) \approx
    \frac{2 e^2 \mu}{\kappa \hbar^2 a}\frac{1}{1 + (q\xi)^{-2}}.
\end{align}
At $\xi^{-1} < q < a^{-1}$ we have
\begin{align}\label{eq:omega_ti_3d}
    \omega_p^2(q) \approx \omega_{p3}^2 = 
    \frac{2 e^2 \mu}{\kappa \hbar^2 a}.
\end{align}
Here $\omega_p$ is independent on $q$, similarly to the 3D plasmon dispersion in Eq.~\eqref{eq:omega_3d0}. This shows the electrostatics at length scale $a<x<\xi$ is similar to the conventional 3D electrostatics with uniform dielectric constant $\kappa$.
At length scale even larger than $\xi$ we have
\begin{align}\label{eq:omega_ti_1d}
    \omega_p^2(q) \approx \frac{e^2 \mu q^2 a}{\kappa_e \hbar^2} \ln (1/qa),\; q < \xi^{-1},
\end{align}
Eq.~\eqref{eq:omega_ti_1d} looks similar as the 1D plasmon dispersion in Eq.~\eqref{eq:omega_1d0} with $\kappa$ replaced by $\kappa_e$.
This shows that the electrostatics at length scale $x>\xi$ returns to the conventional 1D electrostatics with uniform dielectric constant $\kappa_e$.
The crossover of $\omega_p(q)$ between two limits $q \xi \ll 1$ and $q \xi \gg 1$ is again shown in Fig.~\ref{fig:omega}.

The range of $q$ at which Eq.~\eqref{eq:omega_ti0} is valid can be calculated similarly using the criterion Eq.~\eqref{eq:correction0}, where $\mu = \hbar v_F k_F$ for TI.
This leads to
\begin{align}
    q <
\begin{cases}
a^{-1}, \, & k_F a > \alpha^{-1}, \\
 \sqrt{\alpha k_F/a}, \, & 1< k_F a < \alpha^{-1}, 
\end{cases}
\end{align}
where $\alpha = e^2/\kappa \hbar v_F$ is the effective fine structure constant. 
For TI we have $\kappa \sim 200$~\cite{richter1977,borgwardt2016,bomerich2017} and $v_F = 4 \times 10^5$ m/s~\cite{jszhang2011}, so that $\alpha = 0.027$.

Above we discuss the thick wire limit such that electrons occupy many subbands $\mu \gg \hbar v/a$.
Below we discuss the opposite case where electrons only occupy the first subband, and the Fermi level is slightly higher than the first electron subband labeled by $\abs{l} = 1/2$.
Namely, $\mu - \hbar v_F/2a \ll \hbar v_F/a$. 
In this case, one can expand the Dirac energy dispersion Eq.~\eqref{eq:energy_dirac} around $k = 0$ so that
\begin{align}\label{eq:energy_sub}
    \varepsilon_{s\vb{k}} = s\qty[\frac{\hbar v_F}{2a} + \frac{\hbar k^2}{2 m_a} ],
\end{align}
where the effective mass is defined as $m_a = \hbar/2 a v_F$.
For convenience, one can define the Fermi wave vector as $k_F = \sqrt{2m_a (\mu - m_a v_F^2)/\hbar^2}$, such that the condition $\mu - \hbar v_F/2a \ll \hbar v_F/a$ is equivalent to $k_F a \ll 1$.
Since the energy dispersion Eq.~\eqref{eq:energy_sub} near the bottom of the first subband is parabolic, the result of the polarization bubble is similar to a 1D trivial wire Eq.~\eqref{eq:pi_1d1}, as will be shown below.
To see this is the case, first we separate the polarization bubble in Eq.~\eqref{eq:pi_ti} into two parts
\begin{align}
    \Pi(\vb{q},\omega) = \Pi_+(\vb{q},\omega) + \Pi_-(\vb{q},\omega),
\end{align}
where the intraband bubble $\Pi_+(\vb{q},\omega)$ is contributed from terms with the same band indices $s=s'$, and the interband bubble $\Pi_-(\vb{q},\omega)$ is contributed from terms with different band indices $s\neq s'$.
In our case $k_F a \ll 1$, the intraband bubble $\Pi_+(\vb{q},\omega)$ dominates and the interband bubble $\Pi_-(\vb{q},\omega)$ can be neglected, because the interband coupling is strongly suppressed by the mass gap by an extra factor $(\hbar \omega/m_a v_F^2)^2 \ll 1$. 
As a result, in the long wave length limit $q \to 0$ with fixed $\omega$ in the interval $(\hbar q^2/2m_a, \mu/\hbar)$, the polarization bubble reads 
\begin{align}\label{eq:pi_ti2}
    \Re\Pi(q,\omega) \approx \Re\Pi_+(q,\omega) = \frac{n_1}{\pi a^2}\frac{q^2}{m_a\omega^2},
\end{align}
where $n_1 = k_F/\pi$ is the 1D concentration of surface electrons.
Notice Eq.~\eqref{eq:pi_ti2} has the same expression as Eq.~\eqref{eq:pi_1d1} in the trivial wire case by replacing $m^{\star}$ with $m_a$. This happens because both TI and trivial wires have parabolic energy dispersion in the limit $k_F a \ll 1$, which leads to the same polarization bubble.

Substituting Eqs.~\eqref{eq:pi_ti2} and \eqref{eq:vqm} with the second argument $m=0$ into Eq.~\eqref{eq:epsilon} and solving equation $\epsilon(q, \omega_p) = 0$ for $\omega_p$ we arrive at the plasmon dispersion Eq.~\eqref{eq:omega_trivial}. In the limit of large dielectric constant mismatch $\kappa \gg \kappa_e$, the plasmon dispersion $\omega_p(q)$ can be further simplified to arrive at Eq.~\eqref{eq:omega_trivial0}.
This result is justified if $q < \sqrt{\alpha k_F/a}$.

\section{Conclusion}
\label{sec:conclusion}
The dispersion law of plasmons running along thin wires with radius $a$ is known to be practically linear. We show that in wires with a dielectric constant $\kappa$ much larger than that of its environment $\kappa_e$, such dispersion law crosses over to the dispersionless 3D-like law when the plasmon wavelength becomes shorter than the characteristic length $\xi = (a/2) \sqrt{(\kappa/\kappa_e)\ln(\kappa/2\kappa_e)}$ at which the electric field lines of a point charge exists from the wire to its environment. 
This happens both in trivial semiconductor wires and 3D TI wires. 
Our results are related to the electrostatic confinement of the electric field of plasma waves inside a thin wire~\cite{keldysh1997,finkelstein2002,teber2005,kamenev2006,cui2006,huang2021c}. 
They are similar to the previously studied plasmon dispersion law crossover for thin films made by trivial semiconductors~\cite{bondarev2017}, or 3D TIs~\cite{stauber2013}, where the physics is again driven by the electrostatic confinement of the electric field of plasma waves inside the film~\cite{rytova1967,chaplik1971,keldysh1979,huang2021a}.

In this paper, we assumed that the dielectric constant of the wire material $\kappa$ does not depend on $\omega$. This requires to deal with relatively small electron concentrations $n_3$. For example, for STO with electron concentration $n_3=10^{18}$ cm$^{-3}$, using the effective mass $m^{\star}=1.5 m_e$~\cite{Ahrens2007} where $m_e$ is the free electron mass and the room-temperature dielectric constant $\kappa=300$, we arrive at $\hbar\omega_{p3} = 2$ meV, while the soft mode energy $\hbar \omega_s(q=0)$ [which determines the dielectric constant dispersion $\kappa(\omega)$] at room temperature is equal to 11 meV~\cite{Yamada1969}. At larger electron concentrations, for example, for STO with $n_3 > 10^{20}$ cm$^{-3}$ such that $\omega_{p3} > \omega_s(q=0)$, one can include $\kappa(\omega)$ dependence and find the plasmon frequency self-consistently. 
This generalization of our theory is beyond the scope of our paper.

\begin{acknowledgments}
We are grateful to A. Chaplik, M. Entin, M. Fogler, A. McLeod, and B. Skinner for reading the manuscript and useful comments.
Y.H. was partially supported by the William I. Fine Theoretical Physics Institute.
\end{acknowledgments}

%

\end{document}